\documentclass{article}

\PassOptionsToPackage{numbers}{natbib}

\usepackage[preprint]{neurips_2023}
\usepackage{graphicx}

\usepackage[utf8]{inputenc} 
\usepackage[T1]{fontenc}    
\usepackage{hyperref}       
\usepackage{url}            
\usepackage{booktabs}       
\usepackage{amsfonts}       
\usepackage{nicefrac}       
\usepackage{microtype}      
\usepackage{xcolor}         
\usepackage{amsmath}        
\usepackage{bm}
\usepackage{cleveref}      
\usepackage{soul}           
\usepackage{booktabs, multirow} 
\usepackage{soul}
\usepackage{makecell} 
\usepackage{arydshln}

\newcommand{\transformer}{Transformer}
\newcommand\Tstrut{\rule{0pt}{2.6ex}}  
\newcommand{\spann}{\text{span}}
\newcommand{\masked}{\text{masked}}
\newcommand{\mask}{\text{mask}}

\title{Anatomy of Industrial Scale Multilingual ASR}

\author{%
   Francis McCann Ramirez* \quad Luka Chkhetiani* \quad Andrew Ehrenberg \quad Robert McHardy \\ 
  \textbf{Rami Botros} \quad \textbf{Yash Khare} \quad \textbf{Andrea Vanzo} \quad \textbf{Taufiquzzaman Peyash} \quad \textbf{Gabriel Oexle} \\
  \textbf{Michael Liang} \quad \textbf{Ilya Sklyar} \quad \textbf{Enver Fakhan} \quad \textbf{Ahmed Etefy} \quad \textbf{Daniel McCrystal} \\
  \textbf{Sam Flamini} \quad \textbf{Domenic Donato} \quad \textbf{Takuya Yoshioka}
  \\
  AssemblyAI Inc.\\
  \texttt{\url{https://www.assemblyai.com/research/universal-1}} \\
}

\begin{document}

\maketitle

\begin{abstract}
This paper describes AssemblyAI's industrial-scale automatic speech recognition (ASR) system, designed to meet the requirements of large-scale, multilingual ASR serving various application needs. Our system leverages a diverse training dataset comprising unsupervised (12.5M hours), supervised (188k hours), and pseudo-labeled (1.6M hours) data across four languages. We provide a detailed description of our model architecture, consisting of a full-context 600M-parameter Conformer encoder pre-trained with BEST-RQ and an RNN-T decoder fine-tuned jointly with the encoder. Our extensive evaluation demonstrates competitive word error rates (WERs) against larger and more computationally expensive models, such as Whisper large and Canary-1B. Furthermore, our architectural choices yield several key advantages, including an improved code-switching capability, a 5x inference speedup compared to an optimized Whisper baseline, a 30\% reduction in hallucination rate on speech data, and a 90\% reduction in ambient noise compared to Whisper, along with significantly improved time-stamp accuracy.
Throughout this work, we adopt a system-centric approach to analyzing various aspects of fully-fledged ASR models to gain practically relevant insights useful for real-world services operating at scale.
\end{abstract}

\section{Introduction}
\def\thefootnote{*}\footnotetext{These authors contributed equally to this work.}

In the past decade, both the research community and the industry have achieved unprecedented progress in artificial intelligence (AI). Especially after the invention of the \transformer ~\citep{vaswani2017attention}, technical differences among various AI domains, including automatic speech recognition (ASR), computer vision (CV), and natural language processing (NLP), have largely diminished. As a result, the pursuit of higher-quality AI models has become largely synonymous with scaling in terms of model size and the amount of training data~\citep{kaplan20scaling,zhang2022bigssl,dehghani2023scaling}, while exploration of \transformer{} variants, such as Conformer~\citep{gulati2020conformer}, has continued. This principle has been empirically validated across various AI domains. The correlation between data abundance, parameter count, and model performance has been a cornerstone of AI research, driving efforts to amass extensive datasets for training purposes and scale up the training hardware. Importantly, scaling with respect to both the model size and data quantity has resolved issues that smaller models trained on well-organized yet limited datasets have traditionally encountered, including the lack of generalizability.

In the field of ASR, OpenAI's Whisper~\citep{OpenAI-Whisper} has shown the critical impact of data and model scaling. Whisper models are based on a Transformer encoder-decoder architecture, each having a different number of parameters. All models are trained on 680k hours of transcribed speech in multiple languages. The largest model demonstrated industry-leading (at the time of publication of~\citep{OpenAI-Whisper}) English ASR performance across various test sets while exhibiting strong multilingual ASR and translation capabilities. Following this seminal work, multiple papers have been published describing full-fledged ASR systems, including Open Whisper-Style Speech Models (OWSM)~\citep{peng2023reproducing,peng2024owsm}, Universal Speech Model (USM)~\citep{zhang2023google},  SeamlessM4T~\citep{communication2023seamlessm4t}, and Canary-1B~\citep{puvvada2024canary}. Notably, all these models are multilingual by design.

This trend gives rise to the need for a new approach to understanding an ASR system in its entirety. In this approach, one first builds a full-blown ASR system with model architectures and training procedures of interest by using an industry-scale training dataset, and subsequently analyzes the behavior of the system from various viewpoints. This is in contrast to a more traditional reductionist approach where one concentrates on studying a single particular aspect (such as a model architecture or a training algorithm) while normalizing all other variables. Considering that the goal of ASR is to work reliably out of the box under various conditions, we believe the system-centric approach is a valid methodology when one wants to obtain practically relevant insights. It would also shed light on certain aspects that have received less attention thus far, including ASR hallucinations and code-switching, which are traits that Whisper is known to exhibit empirically.

With this perspective in mind, in this paper, we build multilingual ASR models using 12.5M hours of pre-training audio and a 1.8M-hour fine-tuning speech dataset. 
Reflecting our own needs, we focus on a set of high-resource languages---English, Spanish, German, and French.
For model creation, we employ a Conformer encoder~\citep{gulati2020conformer} with a Recurrent Neural Network Transducer (RNN-T) decoder~\citep{graves2013speech} as the model architecture and mostly follow the Google USM paper~\citep{zhang2023google} for training. Through rigorous experimentation and evaluation, we demonstrate that our multilingual models exhibit remarkable robustness and competitive performance with state-of-the-art ASR models such as Whisper large-v3\footnote{\url{https://github.com/openai/whisper/discussions/1762}} and Canary-1B across languages and test sets, while having around half the parameter count. We analyze various aspects of our models, including code-switching, hallucinations, inference latency, timestamp estimation, and the effect of pre-training. Our analysis reveals that ASR models trained on multilingual corpora exhibit an ability of handling code-switching even though the training dataset did not contain code-switching samples. 

This work served as a foundation for building Universal-1, AssemblyAI's commercial ASR system\footnote{\url{https://www.assemblyai.com/blog/announcing-universal-1-speech-recognition-model/}}.
As such, the models and inference code used in this paper are not exactly the same as those deployed in our production system. For ease of reference, we call our models as Universal-1 in this paper.


\section{Related Work}

\subsection{Self-supervised learning and scaling ASR training}

Numerous studies have shown the effectiveness of learning speech representations from speech-only corpora with self-supervised approaches. Since learning to reconstruct audio signals can be challenging, much of the recent research has moved away from auto-encoding approaches \citep{mohamed2022self}. Instead, a speech encoder can be trained to solve proxy pretext tasks, which helps it obtain rich latent speech representations. The encoder can then be further fine-tuned in conjunction with a decoder with an ASR loss. 

Wav2vec 2.0 \citep{baevski2020wav2vec} is trained to solve a contrastive task, where it learns to distinguish between positive and distractor samples, which are obtained after encoding, masking and quantizing different time slices within an utterance. HuBERT \citep{hsu2021hubert} uses a BERT-like \citep{devlin2018bert} masked prediction loss on k-means-clustered audio features. BEST-RQ \citep{chiu2022bestrq}, a method which we adopt in this paper, also uses a masked BERT-style loss. However, it obtains the discrete target labels using frozen random projections and codebooks, thus bypassing the need for elaborate representation learning of discrete prediction targets. These studies and others have shown how pre-training with self-supervised learning (SSL) can reduce the need for large amounts of supervised data.

There are other approaches to scaling the training data, and these studies have also shown that sheer scaling of the training data as well as the model size remains an effective way for improving ASR performance.  
One established approach is 
using pseudo-labels, as in a noisy-student training setting. It was shown that training using pseudo-labels could result in better performance than training with human labels alone~\citep{nst_hwang2022large,nst_hwang2022pseudo}. 
Our previous work \citep{zhang2024conformer1} also showed the benefits of scaling, which we expand upon in this paper with larger models trained on more data. An alternative approach is using weakly supervised data, as with Whisper. 

\subsection{Multilingual ASR}


Thanks to the increased training data quantities and the training algorithms for effectively utilizing them, including SSL, it is becoming possible to train a unified multilingual model encompassing a number of languages~\citep{OpenAI-Whisper,ml_li2022asr2k,ml_barrault2023seamlessm4t,zhang2023google,ml_team2023gemini}. Overall, combining SSL and model scaling has been shown to  simultaneously achieve good performance on hundreds of languages, including low-resource ones. 

Conversely, our work focuses on a specific language family, aiming to study more practical aspects of a production ASR system that go beyond Word Error Rate (WER) evaluation, including timestamp estimation, inference speed, and robustness against hallucination. 

\subsection{Practical aspects of ASR}

End-to-end ASR models are trained to accurately output token sequences, without explicit loss terms for word-level timestamps. Nevertheless, studies have shown that RNN-T and CTC losses, which explore all possible monotonic alignments between the audio and the target transcripts, can learn time alignments as a side effect \citep{zeyer2021does,tian2022bayes,trans_loss_li2019improving,zhao2021addressing}. We experimentally demonstrate that a scaled up bidirectional encoder pre-trained with BEST-RQ and fine-tuned with the RNN-T loss yields accurate word-level timestamp estimates. On the other hand, encoder-decoder models such as Whisper large-v3 and Canary-1B have no notion of blank tokens or alignment-based loss functions. Therefore, additional models are often used to perform forced alignment and obtain the word-level timestamps, as with \citet{zhao2021addressing,bain2022whisperx}.



In a broader AI context, there are various studies investigating the hallucination problem of large language models and other generative models. See, for example, \citet{ji2023survey} for an overview. Recently, this problem has started gaining attention in ASR, becoming increasingly noticeable as ASR models grow in scale \citep{koenecke2024careless, mittal2024towards}. Previous investigations have attempted to illustrate hallucination examples qualitatively. In this paper, we introduce new hallucination metrics to quantitatively analyze the ASR hallucination problem for our RNN-T-based model as well as Whisper and Canary-1B, both of which are based on encoder-decoder models. 

\section{Universal-1}

This section describes the process of developing our multilingual ASR model, Universal-1. The description is split into four parts: training data, model architecture, training method, and inference, each described in Section \ref{sec:training_data}, \ref{sec:model_architecture}, \ref{sec:training_method}, and 
\ref{sec:inference}, respectively. In real usage, additional text formatting is applied for inverse text normalization, punctuation, and truecasing, which we describe in \Cref{app:text_formatting}.

\subsection{Data preparation}
\label{sec:training_data}

To build a robust multilingual model that will perform well under a wide range of conditions, we utilize large, high-quality datasets for training. We employ three types of data, each differing in quantity and reference transcription accuracy levels, as detailed below.
The data quantity of each category for each language is summarized in \Cref{tab:data_counts}. 

\subsubsection{Unsupervised data}
Firstly, we utilize 12.5 million hours of unsupervised, or untranscribed, multilingual audio-only data. This dataset is used for pre-training the encoder of our model, as discussed in \Cref{sec:training_method}. Obtaining unsupervised data in large quantities is relatively straightforward.

Our unsupervised data is acquired from publicly available sources as well as our partners. To ensure that the data contains speech, we employ neural Voice Activity Detection (VAD) from the Silero VAD package~\citep{SileroVAD} to calculate a speech existence ratio for each file. This ratio is defined as the ratio of the speech segment duration to the total audio duration. Any audio files with a speech existence ratio below 70\% are discarded. Additionally, we limit the duration of audio files to be between 8 and 64 seconds. During pre-training, when utilizing this dataset, we randomly select a 32-second segment or apply padding to reach 32 seconds.

\begin{table}[t]
\centering
\caption{Amounts of data (in hours) per language for varying levels of supervision. The table demonstrates the scarcity of supervised data for non-English languages, necessitating the use of pseudo-labeled data.}
\label{tab:data_counts}
\vspace{-.5em}
\begin{tabular}{lccc}
\toprule
Language & Unsupervised & Supervised & Pseudo-labeled \\
\midrule
English & 5,192,686& 149,070 & 1,086,291 \\
Spanish & 1,501,603 & 10,517 & 171,857 \\
German & 1,500,073& 14,696  & 164,737 \\
French & 1,452,665 & 13,993 & 198,263 \\
Others & 2,921,147 & - & - \\ \hline
Total & 12,568,174 & 188,276 & 1,621,148 \\
\bottomrule
\end{tabular}
\end{table}

\subsubsection{Supervised data}

To train an ASR model end-to-end, including a decoder, supervised data with reference transcriptions are indispensable. We have purchased and curated our supervised dataset from various sources, including publicly accessible sources with permissive licensing terms and our partners, among others. Despite the expectation that reference transcriptions are 100\% accurate, we have found that this is not always the case. Therefore, we have developed a filtering scheme for quality control that compares the reference transcriptions against ASR outputs for verification. This quality control pipeline employs several heuristics using various metrics, including WER, consecutive deletion errors, and language detection consistency. After the quality control filtering, our supervised dataset amounts to approximately 188k hours.

\subsubsection{Pseudo-labeled data}
While our supervised dataset may be comparable to or exceed the training data typically used in published studies in terms of quantity, it still does not seem to reflect the variability of real-world speech. To bridge the gap between the quantity of supervised data and the amount ideally required, we also employ a pseudo-labeling approach, in which we generate reference transcriptions with an existing, well-trained ASR model. The effectiveness of pseudo-labeling has been demonstrated previously~\citep{Qizhe20NST, hwang2022pseudo}, including in our past Conformer-1 project~\citep{zhang2024conformer1}.

Since ASR models generate transcription errors, we utilize two ASR models and discard any samples where the two machine transcriptions  disagree to the extent that the WER between them exceeds 20\%. This step helps prevent the model from replicating error patterns of the existing ASR models to some extent. This yields 1.62M hours of pseudo-labeled data. All transcribed samples are segmented to 30 seconds or shorter for training.

\subsection{Model architecture}
\label{sec:model_architecture}

Our ASR model is based on Conformer~\citep{gulati2020conformer}, following the recipe of \citet{zhang2023usm}. 
Our encoder first performs $4\times$ temporal reduction via two convolutional subsampling layers. This is followed by a stack of Conformer layers with a non-streaming configuration using bidirectional attention. We use absolute sinusoidal positional encodings and chunk-wise attention~\citep{zhang2023usm} with a chunk size of $8$ seconds. In most experiments, we use RNN-T~\citep{graves2012rnnt} as a decoder for performing ASR, except for some ablation studies where we also utilize CTC~\citep{graves2006ctc}. 

Our models have approximately 600M parameters, unless noted otherwise. 
Specifically, we employ $24$ Conformer layers with a hidden dimension of $1024$ and $8$ attention heads. Our decoders use a vocabulary size of $2048$ WordPiece~\citep{devlin-etal-2019-bert} tokens by default. The model accepts 80-dimensional log-mel spectrograms as input.



\subsection{Training method}
\label{sec:training_method}

\begin{figure}[ht]
   \centering
   \includegraphics[width=0.7\textwidth]{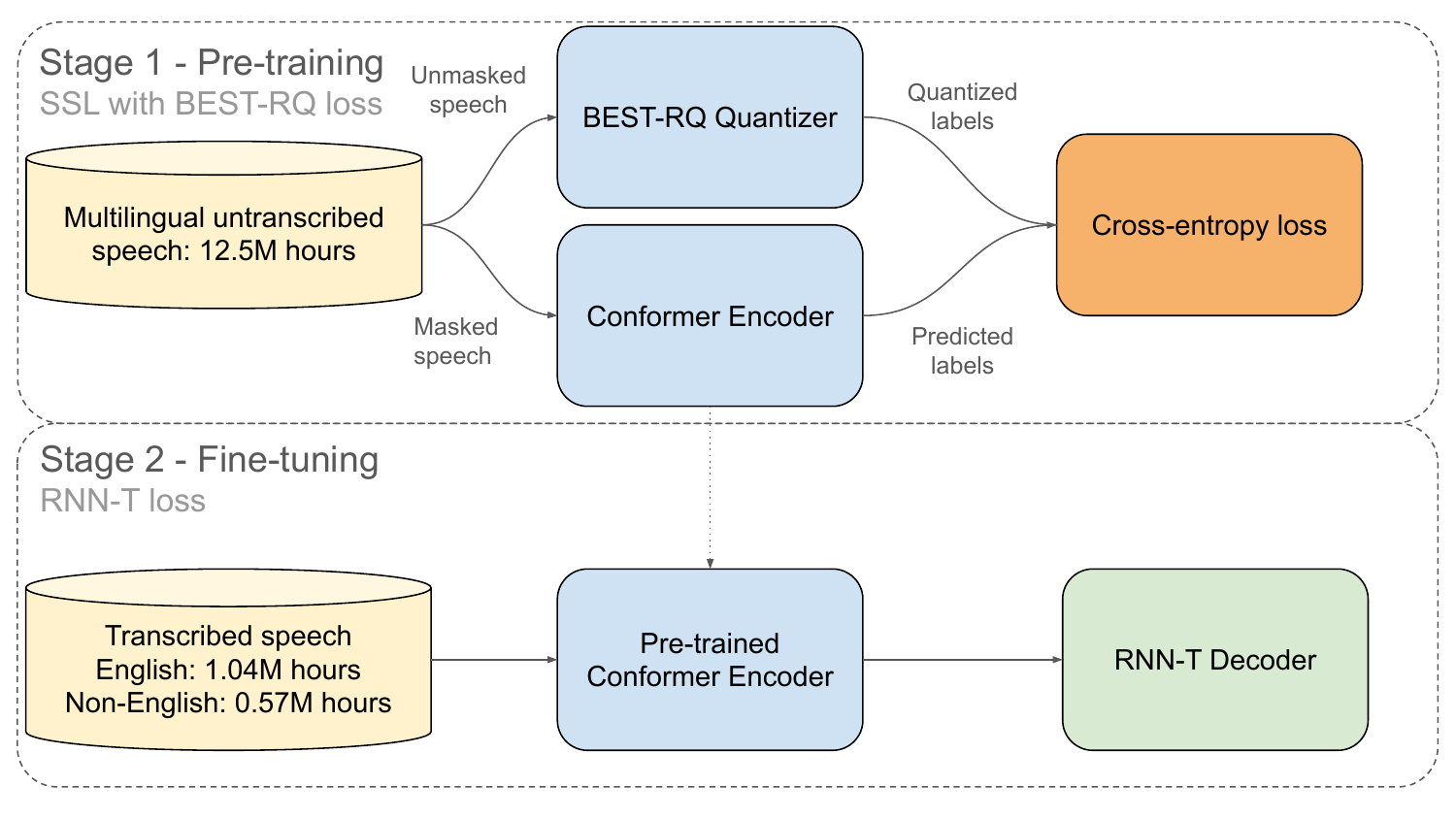} 
   \caption{Two-stage training procedure comprising self-supervised pre-training based on BEST-RQ followed by RNN-T fine-tuning.}
   \label{fig:training-stages}
\end{figure}

We employ a two-stage training procedure, comprising  self-supervised pre-training and fine-tuning. 
\Cref{fig:training-stages} illustrates the overall processing flow of the training procedure. 
The pre-training stage leverages all unlabeled data to pre-condition the encoder. The fine-tuning stage appends a randomly initialized decoder on top of the pre-trained encoder and trains the entire model by using all labeled data, including the pseudo-labeled samples. 

The pre-training and fine-tuning stages are detailed in Sections \ref{sec:pre-training} and \ref{sec:fine-tuning}, respectively, below. Our implementation utilizes JAX~\citep{jax2018github} and Flax~\citep{flax2020github}, and the training is performed on Google Cloud TPU v5e chips using fully-sharded data-parallel (FSDP) parallelism.

\subsubsection{Pre-training}
\label{sec:pre-training}

We utilize BEST-RQ~\citep{chiu2022bestrq}, along with the modifications proposed in \citet{zhang2023usm},
for pre-training Conformer encoders. BEST-RQ offers advantages over other frequently used SSL methods that require additional representation learning, such as Hubert~\citep{hsu2021hubert}, particularly when dealing with large datasets. This advantage is realized by the use of a random projection quantizer and a randomly initialized codebook which remain frozen during training.

Specifically, for a given audio sample, we select a random 32 second window and compute its log-mel spectrogram of shape $(n, D)$, where $n$ and $D$ denote the sequence length and the feature dimensionality, respectively. We use a masking probability, $p_{\mask}$, of $0.01$ to obtain the number of masked regions as $n_{\masked} = n \cdot p_{\mask}$. Then, we use a uniform distribution over the sequence dimension to randomly determine the starting frame of each masked region. For each starting frame, we select a span of $n_{\spann} = 10$ consecutive frames to be masked, allowing for overlaps between neighboring masked regions. This guarantees an upper limit of $n_{\masked} \cdot p_{\mask} \cdot n_{\spann}$ masked frames and introduces variability in the masked frame number and region length. 
The features at the masked frames are replaced with random vectors sampled from $\mathcal{N}(\bm{0}, \sigma \bm{I})$, where $\sigma$ = 0.1, and $\bm{0}$ and $\bm{I}$ are a $D$-dimensional zero vector and an identity matrix, respectively. After this step, the masked representations are fed through the encoder which is followed by a classifier with $Q$ classification heads, where the targets for each of the classification heads are obtained by using a separate BEST-RQ quantizer on the unmasked representations. In our experiments, we use $Q = 8$  classification heads/quantization targets. For efficiency, we only quantize frames which have been masked as other positions do not influence the loss.

Note that the masking method described above slightly differs from the original BEST-RQ algorithm described in \citet{chiu2022bestrq}, where
whether to start a masking region is determined for each frame independently with a fixed masking probability. 
Our implementation guarantees static shapes, enabling more efficient implementation on TPUs. In our preliminary experiments, the two masking strategies showed almost the same convergence behaviors.

\Cref{tab:learning_rates} shows training hyper-parameters for this Section and the next one. Further details about the pre-training process are provided in \Cref{app:pre-training}.

\begin{table}[]
\centering
\caption{Hyperparameters used for pre-training (PT) and fine-tuning (FT).}
\label{tab:learning_rates}
\vspace{-.5em}
\renewcommand{\arraystretch}{1.2}
\begin{tabular}{lccc}
\toprule
             & Encoder PT &  Encoder FT & Decoder FT                     \\
\midrule
Batch size   & 2048                                      &  4096 & 4096                                                           \\ \hline
Peak LR      & 4e-4                                      & 9e-4                         & 3e-3                           \\ \hline
Warm-up steps & 25k                                       & 625                           & 187                            \\ \hline
\multirow{2}{*}{Optimizer}    &  \multicolumn{3}{l}{AdamW, $\beta_1=0.9$, $\beta_2=0.999$, $\epsilon=10^{-8}$,}  \\
& \multicolumn{3}{l}{gradient clipping norm=1.0, weight decay=$10^{-4}$} \\
\bottomrule
\end{tabular}
\renewcommand{\arraystretch}{1}
\end{table}

\subsubsection{Fine-tuning}
\label{sec:fine-tuning}

During the fine-tuning stage, 
the BEST-RQ pre-trained encoder is combined with a randomly-initialized CTC or RNN-T decoder to form an end-to-end ASR model. At the start of fine-tuning, gradients are noisy due to the randomly initialized decoder. To prevent catastrophic forgetting of the pre-trained encoder weights,  we use 
a lower peak learning rate and longer warm-up period for the encoder, as shown in \Cref{tab:learning_rates}. We used float32 precision for fine-tuning, as bfloat16 caused loss spikes. Training is terminated after 75k steps. 
Based on preliminary experiments and educated guesses about the data quality and diversity, we sampled 1.5 times more often from the supervised dataset than from the pseudo-labeled dataset when creating mini-batches. 


\paragraph{Sequential transducer loss}

\begin{figure}[ht]
   \centering
   \includegraphics[width=0.95\textwidth]{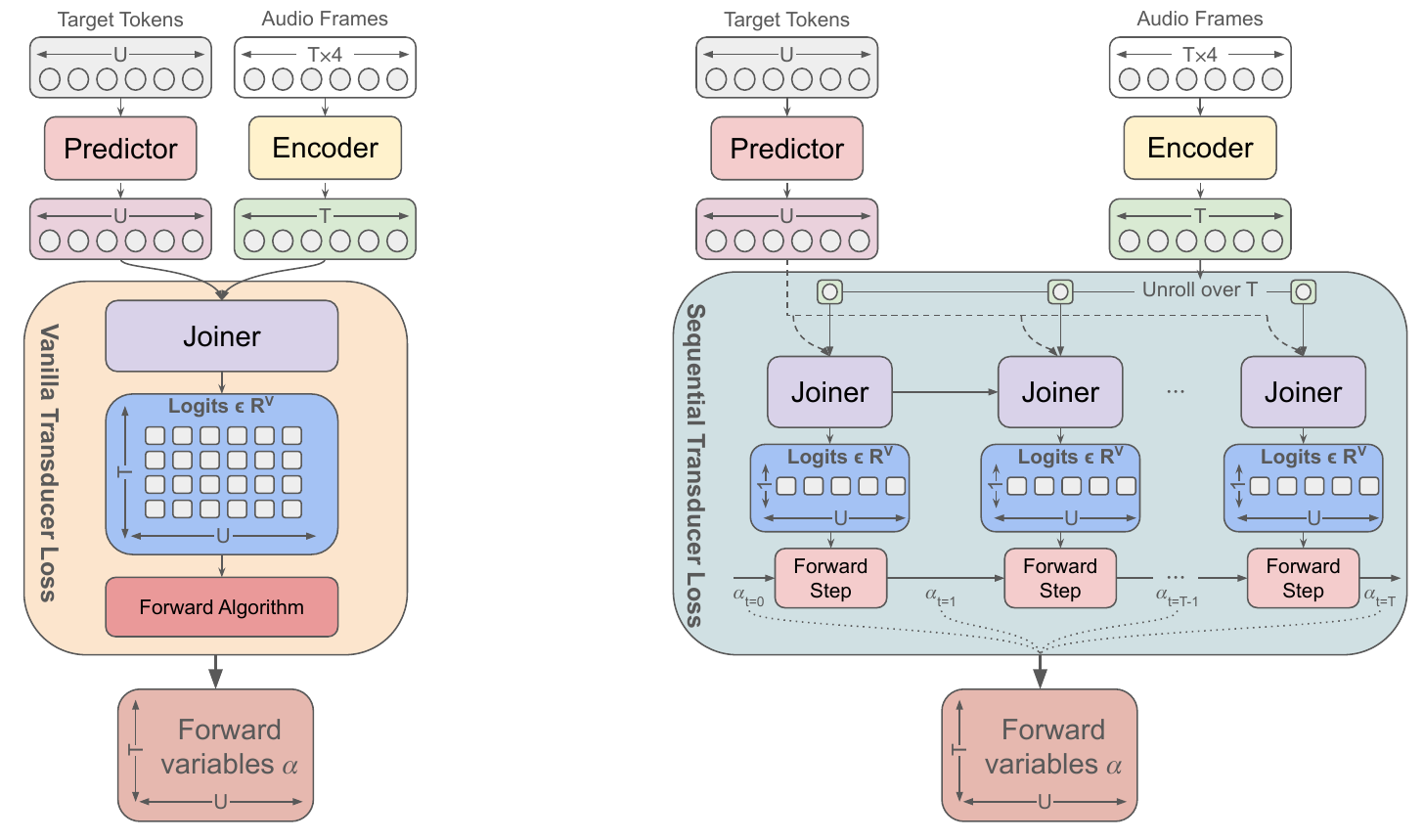} 
   \caption{Overview of the changes made for the sequential transducer loss. Encoder outputs are unrolled over the Time(T) dimension and preventing the creation of a high memory lattice of shape $B \times V \times T \times U$.}
   \label{fig:low-mem}
\end{figure}

While CTC models can easily be trained with high speed on accelerators, training RNN-T models presents challenges when implemented naively. In RNN-T training, the transducer loss is computed over all possible alignments between an encoder output (padded to maximum length $T$) and a target transcript (padded to maximum length $U$)~\citep{graves2012rnnt}. A straightforward implementation, like the one available in PyTorch~\citep{pytorch2017}, would perform the forward-backward algorithm completely in parallel. This requires a fully materialized lattice of output logits of shape $B \times T \times U \times V$, where $B$ and $V$ are the mini-batch and vocabulary sizes, respectively. In our case, this lattice alone would require $4096 \cdot 800 \cdot 256 \cdot 2048 \cdot 32 \text{bits} = 6.9 \text{TB}$ of TPU memory, which is infeasible. This problem might be alleviated by computing the loss separately for each sample, as proposed in \citet{trans_loss_li2019improving,trans_loss_braun2023neural}. 

In our work, we compute the loss sequentially over time steps $t$ instead 
to avoid negatively impacting JAX's automated FSDP sharding. Our algorithm scans over the encoder output for each $t$, computes the joiner outputs and the forward variables $\alpha_t$, which are carried forward as states for the next step. Since XLA would still automatically parallelize the computation of the gradients, we also use a custom gradient function to enforce sequential backpropagation through time (BPTT), passing the gradients from each step to the one before. This is functionally equivalent to computing the backward algorithm sequentially, although we leverage autodiff instead of explicit terms for $\beta_t$. By itself, this sequentialization would slow down the loss computation compared to working on a fully materialized logit lattice. However, since it reduces the memory requirement by a factor of $T$, it allows us to fit a larger batch size, leading to more parallelization through the Conformer encoder and thus achieving a higher training throughput overall for a given amount of available TPU memory. We also unroll this scan loop for 50 time-steps, to get some parallelization. Other advanced approaches for transducer loss computation, such as  \citet{trans_loss_bagby2018efficient,trans_loss_sim2017improving}, are not attempted in this paper.

\subsection{Inference}
\label{sec:inference}

In this section, we outline our inference method for dealing with long-form audio input. At a high level, our inference method consists of three stages:
\begin{enumerate}
    \item \textbf{Audio segmentation}: Segment the input audio of arbitrary size into shorter chunks.
    \item \textbf{Decoding}: Batch forward each chunk through the model and perform greedy decoding to obtain output text and word-level timestamps.
    \item \textbf{Merging}: Merge the chunk-wise outputs and readjust time stamps to the original, pre-segmentation time scale.
\end{enumerate}
Unlike the sequential decoding scheme employed by \citet{OpenAI-Whisper,peng2023reproducing,peng2024owsm}, this divide-and-conquer approach allows batched processing of long-form audio, significantly reducing inference latency.

\subsubsection{Audio segmentation}
Long-form audio input is usually split into shorter chunks. This is performed to reduce the mismatch between the audio length to be decoded and those encountered by the model during training, where the training-time audio length is typically in the order of tens of seconds. Segmentation also helps reduce inference time when the computational complexity of the forward pass grows super-linearly, which is the case with vanilla Transformer models. For instance, Whisper and its follow-up models~\citep{OpenAI-Whisper,peng2023reproducing,peng2024owsm} sequentially transcribe 30-second audio chunks from the beginning of the input audio by shifting the 30-second window based on predicted timestamps.

Our approach is to process individual audio chunks in parallel to optimize for reducing inference latency, i.e., turnaround time, for generating the output transcription for the entire audio. We split the input audio in a way that makes each chunk as long as possible within the maximum input length the model has been trained for during both the pre-training and fine-tuning stages, i.e., 32 seconds. We implement this by leveraging voice activity detection (VAD) to prevent the chunk boundary from occurring in the middle of a word. Specifically, a new chunk is created when a predefined minimum length is exceeded, and the VAD detects consecutive silence of 0.1 seconds or longer. The produced audio chunks are finally converted to log-mel spectrograms, padded to 32 seconds, and batched together up to a predefined maximum batch size. For VAD, we utilize WebRTC~\cite{WebRTC}. 




\subsubsection{Decoding}
The batched audio frames from the segmentation stage are passed through the Conformer encoder of our model to obtain the temporally downsampled, encoded feature frames. These are passed to a batched greedy RNN-T decoding routine, which iteratively generates the output tokens and token-level timestamps for the entire batch in parallel. We limit the decoder to emit up to 5 tokens per feature frame. The timestamps produced at this stage are with respect to each chunk's local reference frame, i.e., in the range from 0 to 32 seconds.

\subsubsection{Merging}
Finally, we take the output transcriptions and timestamps from all batches and concatenate them to generate a single output. This involves concatenating the individual transcriptions and applying per-chunk offsets to the predicted timestamps. That is, we convert the timestamp values from a local (i.e., within-chunk) time scale to a global one. Additional post-processing steps are applied, including timestamp bias correction (\Cref{sec:timestamp_estimation}) and text formatting (\Cref{app:text_formatting}).

\section{Experimental  Results}

\subsection{Evaluation datasets}

As \citet{OpenAI-Whisper} put it succinctly, the goal of ASR should be to develop a single robust model that works reliably without the need for dataset-specific fine-tuning, which they call a ``zero-shot''
setting. Following this approach, we test our ASR models with a variety of test sets for both English and non-English languages by using a mix of publicly available datasets and independently sourced datasets representative of common ASR use cases.

For each test set, we calculate a micro-average word error rate (WER) as a performance evaluation metric. As an overall performance metric, we calculate the macro-average of the WERs of individual test sets. The full list of our test sets is shown in \Cref{app:eval_data}.


\subsection{English ASR}
\label{sec:english_asr}

\Cref{tab:performance_comparison} shows the WERs of our model, two representative open-source ASR models, and four commercial ASR providers. The open-source ASR models used are Whisper large-v3, the latest model of Whisper, and Canary-1B, the most recent powerful open-source model. Both models are based on an encoder-decoder model architecture. Whisper large-v3 consists of 1.55B parameters, with multilingual ASR and X-to-English translation capabilities. Canary-1B has 1B parameters and is trained to perform ASR and translation for English, Spanish, German, and French. For long-form ASR with Canary-1B, we utilized NVIDIA NeMo's chunked inference code\footnote{\url{https://github.com/NVIDIA/NeMo/blob/aee120aaa350c7e9709e6c48092d616f42ebc5cd/examples/asr/speech_multitask/speech_to_text_aed_chunked_infer.py}}. All commercial ASR providers compared are queried using the default settings of their APIs as of April 1st, 2024.

The results demonstrate that, despite having only 600M parameters, our model performs comparably to the much bigger open-source models and outperforms other ASR providers. Unlike other systems, our model remains competitive across all test sets, showing its robustness. Our model performs particularly well for Noisy, consisting of noisy long-form real-world samples, outperforming the next best system by 7.6\% relative.


\begin{table}[t]
    \centering
    \caption{WER (\%) of English ASR: Lower is better; best results highlighted in \textbf{bold}. CV: CommonVoice, LS: LibriSpeech. Podcast, Broadcast, Telephony, and Noisy are our own internally sourced test sets.}
    \vspace{-.5em}
    \label{tab:performance_comparison}
    \begin{tabular}{lccccccc} 
        \toprule
        Dataset & \thead{Universal-1} & \thead{Canary-1B} & \thead{Whisper\\Large-v3} & \thead{Azure\\Batch v3.1} & \thead{Deepgram \\ Nova-2} & \thead{AWS} & \thead{Google \\ Latest-long} \\
        \midrule
        CV V5.1 & \textbf{7.8} & 6.9 & 8.7 & 9.2 & 11.9 & 9.0 & 16.9 \\
        CORAAL & 12.9 & \textbf{9.4} & 12.5 & 12.7 & 10.6 & 11.3 & 19.1 \\
        Earnings-21 & 9.8 & 11.2 & 9.6 & \textbf{7.4} & 12.3 & 10.5 & 11.9 \\
        LS test-clean & 1.8 & \textbf{1.5} & 1.8 & 2.8 & 2.6 & 2.9 & 5.8 \\
        LS test-other & 3.6 & \textbf{3.0} & 3.6 & 6.4 & 5.7 & 6.6 & 12.6 \\
        Meanwhile & \textbf{5.2} & 6.0 & 9.7 & 6.2 & 6.4 & 7.3 & 10.0 \\
        TED-LIUM 3 & 7.6 & 7.8 & 7.4 & 9.7 & \textbf{6.6} & 9.1 & 11.3 \\
        Podcast & \textbf{8.9} & 11.7 & 10.0 & 9.7 & 11.8 & 10.2 & 11.9 \\
        Broadcast & \textbf{4.5} & 5.4 & 4.8 & 6.0 & 6.4 & 5.9 & 8.2 \\
        Telephony & \textbf{10.7} & 13.1 & 12.8 & 16.4 & 15.1 & 16.0 & 20.4 \\
        Noisy & \textbf{10.9} & 12.9 & 11.8 & 16.8 & 15.7 & 27.5 & 21.3 \\ \hline        
        \Tstrut Average & \textbf{7.6} & 8.1 & 8.4 & 9.4 & 9.5 & 10.6 & 13.6 \\
        \bottomrule
    \end{tabular}
\end{table}

\subsection{Multilingual ASR}


\begin{table}[t]
    \centering
    \caption{WER (\%) of multilingual ASR: Lower is better; best results highlighted in \textbf{bold}. LS: LibriSpeech. Private ES/DE/FR are our internally sourced test sets.}
    \label{tab:performance_comparison_languages}
    \vspace{-.5em}
    \begin{tabular}{@{}l*{7}{c}@{}}
        \toprule
        Dataset & \thead{Universal-1} & \thead{Canary-1B} & \thead{Whisper \\ Large-v3} & \thead{Azure\\Batch v3.1} & \thead{Deepgram \\ Nova-2} & \thead{AWS} & \thead{Google \\ Latest-long} \\
        \midrule
        \multicolumn{8}{c}{Spanish} \\ \cline{2-8}
        \Tstrut Fleurs & 4.9 & 7.1 & \textbf{2.8} & 4.9 & 7.0 & 6.2 & 5.0 \\
        Multilingual LS & 3.4 & \textbf{3.0} & 5.7 & 5.7 & 5.8 & 3.4 & 7.2 \\
        Private ES & \textbf{4.3} & 11.3 & 6.6 & 8.9 & 8.2 & 6.2 & 13.7 \\
        Voxpopuli & 7.7 & \textbf{7.1} & 9.6 & 8.8 & 9.3 & 8.6 & 10.7 \\
        CommonVoice V9 & \textbf{3.6} & 4.2 & 5.0 & 7.5 & 8.0 & 4.7 & 7.1 \\ \cline{2-8}
        \Tstrut Average & \textbf{4.8} & 6.5 & 6.0 & 7.2 & 7.6 & 5.8 & 8.7 \\
        \midrule
        \multicolumn{8}{c}{German} \\ \cline{2-8}
        \Tstrut Fleurs & 10.2 & 9.2 & 7.5 & \textbf{7.2} & 12.4 & 9.0 & 12.1 \\
        Multilingual LS & 4.3 & 4.5 & 7.4 & 5.7 & 8.2 & \textbf{3.2} & 12.0 \\
        Private DE & \textbf{7.5} & 11.9 & 8.7 & 9.5 & 11.1 & 9.7 & 12.1 \\
        Voxpopuli & 13.8 & \textbf{9.5} & 12.6 & 14.7 & 15.1 & 16.8 & 17.4 \\
        CommonVoice V9 & 4.7 & \textbf{4.6} & 6.0 & 7.4 & 9.4 & 5.8 & 10.1 \\ \cline{2-8}
        \Tstrut Average & 8.1 & \textbf{7.9} & 8.4 & 8.9 & 11.2 & 8.9 & 12.7 \\
        \midrule
        \multicolumn{8}{c}{French} \\ \cline{2-8}
        \Tstrut Fleurs & 6.9 & 7.7 & \textbf{5.6} & 8.4 & 9.7 & 8.5 & 12.4 \\
        Multilingual LS & \textbf{3.2} & 4.0 & 8.1 & 9.7 & 5.2 & 4.5 & 15.2 \\
        Private FR & 16.8 & 27.6 & 16.1 & 23.7 & 17.5 & \textbf{14.9} & 17.6 \\
        Voxpopuli & 9.5 & 10.1 & 11.2 & 11.4 & 11.2 & \textbf{8.7} & 14.7 \\
        CommonVoice V9 & 8.0 & \textbf{6.5} & 11.3 & 12.9 & 12.1 & 7.4 & 17.5 \\ \cline{2-8}
        \Tstrut Average & 8.9 & 11.2 & 10.5 & 13.2 & 11.1 & \textbf{8.8} & 15.5 \\
        \bottomrule
    \end{tabular}
\end{table}

\Cref{tab:performance_comparison_languages} depicts the WERs of our model and the same six ASR systems, including two open-source models and four commercial ASR services, for Spanish, German, and French test sets. Our model achieved the lowest WER on average. Furthermore, our model was competitive even in test sets for which it did not achieve the best performance, despite its modest model size, which is beneficial for processing a large number of requests with low latency. We believe the robustness and consistently competitive performance across all test sets and languages achieved by our model result from the use of diverse training data and meticulous data filtering.

We would like to caution that there may have been data leakage for some of the considered systems for the FLEURS corpus. For FLEURS, we observed some instances of incorrect reference transcriptions being correctly predicted by some models. 



\subsection{Code-switching}

One emergent capability in multilingual ASR models is the ability to handle code-switching speech. Code-switching occurs when a speaker switches back and forth between two or more languages in a single conversation, sometimes even within a single sentence. We have observed that our model, trained on multilingual training data using a multilingual tokenizer, exhibits the ability to handle code-switching speech to a certain extent. In this section, we explore this aspect by comparing our model with Whisper and Canary-1B, which also use multilingual tokenizers and are trained on multilingual data.

To test the code-switching capability, we created datasets by combining audio clips from LibriSpeech test-clean and MLS corpora. They were chosen since the monolingual ASR accuracy for these datasets is sufficiently high to the extent that allows us to largely isolate the WER degradation caused by handling code-switching. We created 250 code-switching test samples per language pair as follows.


\begin{enumerate}
    \item Randomly determine a target duration from the range between 30 and 180 seconds. 
    \item Randomly select a file from either MLS or LibriSpeech test-clean datasets. 
    \item Continue selecting files  from these two datasets in an alternating fashion until the total audio duration exceeds the target length.  
    \item Concatenate all the selected audio files to generate a single audio file. The reference transcription is also created by concatenating all the corresponding reference transcriptions. 
\end{enumerate}
The duration range of 30--180 seconds in the first step was chosen to provide sufficient coverage with respect to the frequency of language alternations and the dominance of one language over the other in the concatenated files. Each test audio file created in this manner consisted of 3--19 files taken from the LibriSpeech and MLS corpora.

We considered three language pairs: English-Spanish, English-French, and English-German, taking into account the fact that code-switching often occurs between English and another language in most real scenarios. The corresponding test sets are referred to as en-es, en-fr, and en-de.

\begin{figure}[t]
   \centering
   \includegraphics[width=1.0\textwidth]{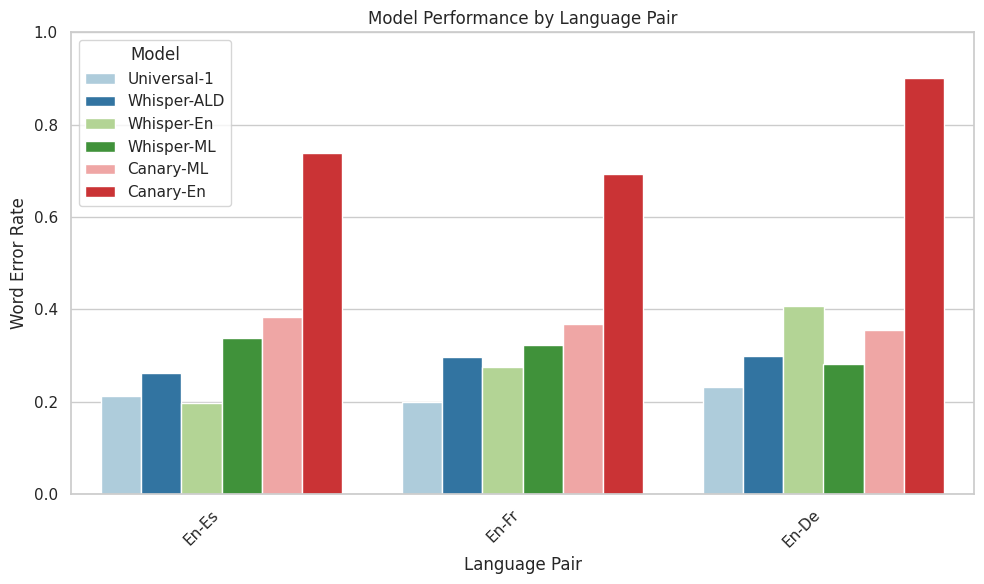} 
   \caption{Code-switching experiment results using synthetic datasets, comparing our model and open-source models. The tags in the legend correspond to different ways of configuring the open-source models. ALD: Whisper's automatic language detection was used to predict the language token for each sample. EN: English was specified. ML: The non-English language of each dataset was specified.}
   \label{fig:cs}
\end{figure}


Unlike our model, which does not enforce a specific language during decoding, Whisper and Canary-1B require a user to specify a language by default. Therefore, for these models, we considered the following conditions:
\begin{enumerate}
    \item Matched language (ml): For each dataset, we specify the language token of the non-English language.
    \item English (en): For each dataset, we specify the English language token. 
    \item Automatic Language Detection (ALD): For Whisper, we let the model predict the language token for each file independently. Canary-1B does not provide this option.  
\end{enumerate}

\Cref{fig:cs} shows the WERs obtained by the three models for each language pair. We can observe that our model handled the code-switching samples quite effectively compared to the other two models, irrespective of their language-code configurations, achieving the lowest WER by substantial margins. Upon manual inspection, we found that other models occasionally introduced disruptive artifacts, which seem to be the result of conditioning decoding on one language token or training for multiple tasks. Sometimes, these artifacts manifest themselves as significant deletion errors or translations, even though we specified the task as transcription. We hypothesize that the stability of our model with respect to code-switching can be attributed to the balanced diversity of our training datasets and the single-task nature of our training setup.


This benchmark was our first foray into quantifying code-switching, and we leave this open as an area for further work. Possible areas to explore include training on data created in a fashion similar to the benchmark we used and studying the trade-off between the use of explicit language tokens and code-switching performance.

\subsection{Inference latency}

To investigate inference latency and throughput, we compared our Universal-1 model with Whisper large-v3 and Canary-1B. The latency of an ASR system is quantified using Real Time Factor (RTF), defined as the ratio of the aggregate wall clock time to process an entire dataset to the cumulative duration of the audio files within the dataset. In addition to the overall RTF analysis, we include detailed measurements of the inference times for both the encoder and decoder components. We utilized two distinct datasets, LibriSpeech test-other and TED-LIUM3, which serve as proxies for short-form and long-form audio benchmarking, respectively. The evaluation was conducted on an Nvidia T4 Graphics Processing Unit (GPU), equipped with 16 Gigabytes (GB) of Video Random Access Memory (VRAM).

In the long-form ASR benchmarking, we enabled batched inference wherever possible. As described in Section \ref{sec:inference}, our model splits a long-form audio input signal and batch-processes the individual chunks. The official implementation of Whisper does not support batched inference. In our experiments, we utilized the batched inference method as implemented in \citet{bain2022whisperx}, which also uses Faster-Whisper for efficient computation. For Canary-1B inference, we employed  the same chunked inference script that was used in the  accuracy benchmarking, as described in Section \ref{sec:english_asr}. Since this implementation does not seem to take advantage of parallel batched computation, we only considered a single-sample batch computation for Canary-1B. For Whisper large-v3 and Universal-1, the maximum possible batch size within memory bounds was used for each model.

Table \ref{tab:latency_benchmarking} presents the latency benchmarking experimental results. For the short-form audio clips, our model achieved 43\% and 60\% relative speed-up compared to Whisper large-v3 and Canary-1B, respectively. These speed gains can be attributed to the smaller parameter count, the use of chunk-wise attention in the encoder, and the relative compactness of a Transducer decoder compared to \transformer{} decoders. 

The inference efficiency of Universal-1 is further enhanced when transcribing long-form audio content. Even without batched inference, we observed  68\% and 87\% relative RTF reduction compared to Whisper large-v3 and Canary-1B, respectively. With batched inference, our model achieved a 5-fold RTF reduction in comparison to Whisper large-v3. The substantial RTF gain can be primarily attributed to the improved decoder speed, which greatly benefited from the parallelized inference. Notably, we observed an increase in encoder RTF with batched inference for both the Whisper large-v3 and Universal-1 models.  We attribute this phenomenon to a computational bottleneck with the the T4 GPU hardware used in this experiment. 

\begin{table}[!htp]\centering
    \caption{Inference latency of Universal-1, Whisper large-v3, and Canary-1B on short-form (LibriSpeech test-other) and long-form (TED-LIUM3) audio. Latency is measured in terms of Real Time Factor (RTF). Lower is better; best results are highlighted in \textbf{bold}. For decoding short-form audio, batch size was always set to 1. For decoding long-form audio, the maximum batch size that fit into memory of the Nvidia T4 GPU was selected for each model.}
    \vspace{-.5em}
    \label{tab:latency_benchmarking}
    \resizebox{\textwidth}{!}{
        \begin{tabular}{lccccccccc}\toprule
        \multirow{3}{*}{Model} &\multirow{3}{*}{\begin{tabular}{@{}c@{}}Model \\ size\end{tabular}} &\multicolumn{3}{c}{LibriSpeech test-other} &\multicolumn{4}{c}{TED-LIUM3} \\\cmidrule{3-9}
        & &\multicolumn{3}{c}{RTF ($\times 10^{-3}$) } &\multirow{2}{*}{\begin{tabular}{@{}c@{}}Batch \\ size\end{tabular}} &\multicolumn{3}{c}{RTF ($\times 10^{-3}$)} \\\cmidrule{3-5}\cmidrule{7-9}
        & &Encoder &Decoder &Total & &Encoder &Decoder &Total \\\midrule
        Canary-1B &1B &14.8 &135.1 &149.9 &1 &6.7 &142.9 &149.6 \\\hline
        \multirow{2}{*}{Whisper large-v3} &\multirow{2}{*}{1.55B} &\multirow{2}{*}{43.3} &\multirow{2}{*}{61.0} &\multirow{2}{*}{104.3} &1 &9.7 &50.2 &59.9 \\
        & & & & &24 &10.0 &19.8 &29.7 \\\hline
        \multirow{2}{*}{Universal-1} &\multirow{2}{*}{600M} &\multirow{2}{*}{\textbf{9.4}} &\multirow{2}{*}{\textbf{50.6}} &\multirow{2}{*}{\textbf{60.0}} &1 &\textbf{2.3} &16.6 &18.9 \\
        & & & & &64 &3.2 &\textbf{2.5} &\textbf{5.7} \\
        \bottomrule
        \end{tabular}
    }
\end{table}


\subsection{Hallucination analysis}

\subsubsection{Definition}
Due to the ongoing amalgamation of deep learning (DL) methods for audio and text processing, the phenomenon of hallucinations, originally discovered in NLP tasks, has become increasingly more common in the ASR field. Anecdotally, the Whisper large-v3 model has particularly been demonstrated to suffer from an increased propensity for hallucinations compared to its predecessors, despite substantial improvements in WER\footnote{\href{https://github.com/openai/whisper/discussions/1762\#discussioncomment-7557044}{https://github.com/openai/whisper/discussions/1762\#discussioncomment-7557044}}. However, there have been limited attempts thus far to study this emerging behaviour of modern ASR systems quantitatively. In this section, we propose evaluation metrics for quantitative analysis of ASR models with respect to their susceptibility to hallucinations.

In the context of ASR, hallucinations occur when the ASR model continuously produces words that are not grounded in the input audio or when it disregards long spoken segments. They can be measured by the number of consecutive transcription errors that the model makes in comparison to the reference transcripts. Following the edit operation concept in edit distance, such errors can be further divided into subgroups such as \textit{fabrication} errors consisting of consecutive insertions or substitutions, and \textit{omission} errors consisting of consecutive deletions. Based on this classification of consecutive errors, we define the corresponding consecutive error rates for a detailed analysis of hallucinations as follows:

\begin{itemize}
  \item Fabrication rate (${F\!R}_{N}$): number of $N$ or more consecutive insertion or substitution errors observed per hour.
  \item Omission rate (${O\!R}_{N}$): number of $N$ or more consecutive deletion errors observed per hour.
  \item Hallucination rate (${H\!R}_{N}$): number of $N$ or more consecutive insertion, substitution or deletions errors observed per hour.
\end{itemize}

\subsubsection{Results on speech audio}

\begin{figure}[ht]
    \centering
    \includegraphics[width=.9\textwidth]{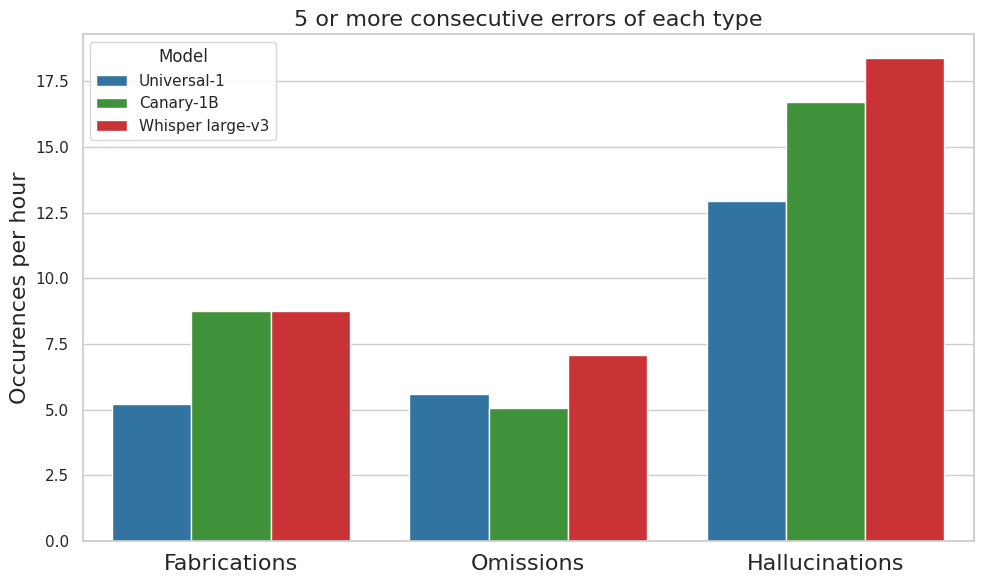}\hfill
    \caption{Occurrences of five or more consecutive errors of each type per hour for Universal-1, Canary-1B and Whisper large-v3 models.}
    \label{fig:absolute_hallucination_rates}
\end{figure}

\begin{figure}[ht]
    \centering
    \includegraphics[width=.9\textwidth]{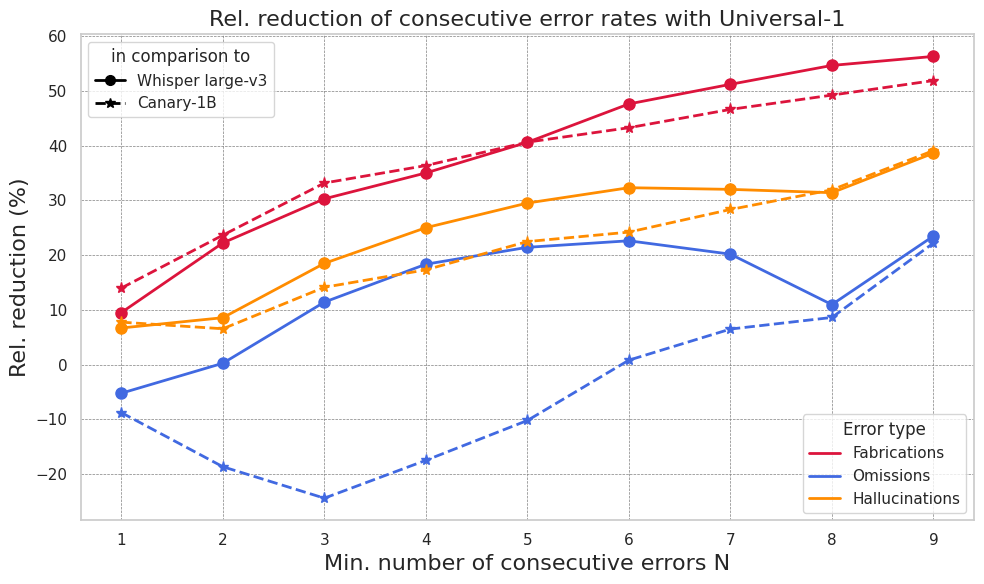}\hfill
    \caption{Rel. reduction of $N$ or more consecutive errors of each type per hour of Universal-1 in comparison to Whisper large-v3 and Canary-1B models for different $N$s.}
    \label{fig:relative_hallucination_rates}
\end{figure}

\Cref{fig:absolute_hallucination_rates,fig:relative_hallucination_rates} show our hallucination analysis results obtained from 146 hours of audio from a diverse set of English datasets\footnote{The datasets used largely overlap with our English test sets listed in Appendix \ref{app:eval_data}.}. We compared our Universal-1 model, Whisper large-v3, and Canary-1B. \Cref{fig:absolute_hallucination_rates} shows the rates of fabrications, omissions, and hallucinations with 5+ consecutive errors, i.e., ${F\!R}_5$, ${O\!R}_5$, and ${H\!R}_5$, respectively. Both Whisper large-v3 and Canary-1B turned out to be similarly susceptible to compounding insertion and substitution errors spanning 5 or more words. On the other hand, our model was much more robust to this type of error, showing a 41\% relative reduction in ${F\!R}_5$. Additionally, it achieved a 21\% relative reduction in ${O\!R}_5$ compared to Whisper large-v3, while being 10\% more prone to consecutive deletions relative to Canary-1B. Overall, we obtained 30\% and 22\% relative reductions in ${H\!R}_5$ compared to Whisper large-v3 and Canary-1B, respectively.


In \Cref{fig:relative_hallucination_rates}, we vary the consecutive error span, $N$, from 1 to 9. For each $N$ value, we show the relative reduction rates in ${F\!R}_N$, ${O\!R}_N$, and ${H\!R}_N$ achieved by Universal-1 in comparison to Whisper large-v3 and Canary-1B. For fabrication and hallucination errors, we observed consistent increases in the relative reduction rates with increasing $N$. For very long consecutive errors spanning 9 or more words, Universal-1 achieved a 50+\% relative reduction in the fabrication rate and an $\approx\!40\%$  relative reduction in the hallucination rate over Whisper large-v3 and Canary-1B. Interestingly, while Canary-1B was more resilient to omissions than Universal-1 for shorter consecutive errors, this trend reversed for $N \geq 6$. As a result, we observed a 20+\% relative reduction in the omission rate for 9 or more consecutive deletions in comparison to both Whisper large-v3 and Canary-1B.


We would attribute the reduced propensity for hallucinations of Universal-1 to the use of a Transducer-based model architecture, as opposed to an encoder-decoder architecture employed by Whisper and Canary-1B, and to our extensive data filtering pipelines for removing training samples with low-quality reference transcriptions. As discussed before in \cite{Hannun2019}, discriminative sequence models have different levels of susceptibility to the label bias problem, which is characterized by the overreliance on previously emitted labels while disregarding new evidence. We hypothesize that the hallucination problem in ASR is partly caused by label bias. Thus, RNN-T could be less susceptible to the aforementioned problem, since its auto-regressive decoder is smaller than in encoder-decoder models.

\subsubsection{Fabrication during ambient noise}
We further analyzed the behaviors of these models on ambient noise data. ASR systems are supposed to produce empty text when applied to ambient noise. Therefore, we focus on the fabrications generated by the ASR models.

We utilized the AudioSet~\citep{hall_gemmeke2017audio} and the DNC~\citep{hall_rafael2021dnc} datasets. Specifically, we obtained 200 audio samples from each dataset, while excluding any sound categories that might contain human speech. The filtering was performed based on the ontology provided by each dataset, removing categories like \emph{singing} or \emph{conversation}. As shown in Table \ref{tab:fabrication_ambient_noise}, Whisper large-v3 and Canary-1B, both based on an encoder-decoder architecture, almost always generated some text for these samples. In comparison, our model produced a blank output around 90\% of the time. For Canary-1B, most of the generated text consisted of short words like ``and'', while Whisper would often output strings of repeated or unintelligible special symbols. The most severe cases---and the hardest ones to filter out with post-processing---are those where the models output long plausible sequences, which Whisper large-v3 suffered from most frequently. Our model's faulty outputs tended to be short, as shown in the table.


While a separate VAD method may be applied to remove non-speech segments from the input audio beforehand, aggressive VAD could result in increased omissions.
Therefore, any lack of robustness against ambient noise could still pose a challenge in practice.

\begin{table}[]
\centering
\caption{Fabrication comparison of three ASR models on ambient noise. Best results are highlighted in bold.}
\label{tab:fabrication_ambient_noise}
\vspace{-.5em}
\renewcommand{\arraystretch}{1.1}
\begin{tabular}{lllccc}

\toprule

 &  & \multirow{2}{*}{Dataset} & \multicolumn{3}{c}{Model} \\
 &  &  & Whisper large-v3 & Canary-1B & Universal-1 \\

\midrule

\multirow{2}{*}{Non-blank response rate} &  & AudioSet & 100\% & 100\% & \textbf{10.5\%} \\

 &  & DNC & 99.5\% & 100\% & \textbf{10.0\%} \\

\hline
 
\multirow{6}{*}{\begin{tabular}[c]{@{}l@{}} Number of characters \\ in non-blank responses\end{tabular}} & \multirow{2}{*}{Mean} & AudioSet & 16 & 5.0 & \textbf{3.6} \\
 &  & DNC & 7.6 & 87.0 & \textbf{2.0} \\

\cline{2-6}

 & \multirow{2}{*}{Median} & AudioSet & 3 & 3 & \textbf{2} \\
 &  & DNC & 3 & 3 & \textbf{2} \\

\cline{2-6}

& \multirow{2}{*}{$\geq 10$} & AudioSet & 35\% & 1.5\% & \textbf{0.5\%} \\
 &  & DNC & 39.5\% & 14\% & \textbf{0\%} \\

\bottomrule
 
\end{tabular}

\renewcommand{\arraystretch}{1}
\end{table}

\subsection{Timestamp estimation} \label{sec:timestamp_estimation}

Many real-world scenarios require accurate word-level timestamps in addition to high-quality transcriptions to build applications with ASR systems. In this section, we evaluate the word-level timestamp performance of our Universal-1 model and several other open-source models.

For the timestamp prediction accuracy evaluation, we used a diverse test set, as listed in Appendix \ref{app:eval_data_timestamps}. To create reference word-level timestamps for the test audio files, we performed forced alignment with the Montreal Forced Aligner toolkit~\citep{mcauliffe17_interspeech} on the audio signal and the reference transcription. For each tested model, the ASR output was first aligned with the reference transcription. We calculated the difference between the reference and predicted timestamps for all words that were matched successfully between the reference and ASR transcriptions.


\begin{figure}[ht]
   \centering
   \includegraphics[width=0.5\textwidth]{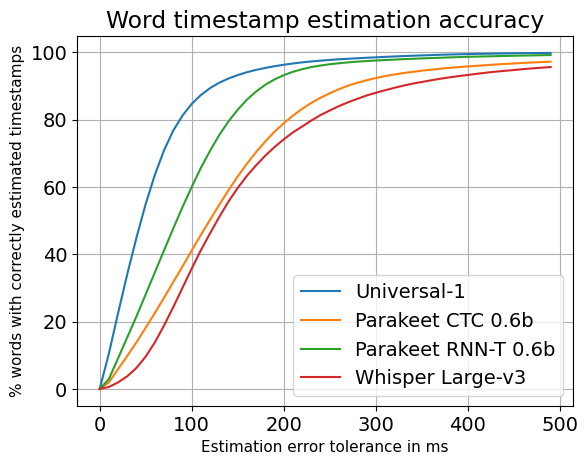} 
   \caption{Word-level timestamp estimation accuracy for different models as a function of estimation error tolerance. A curve closer to the upper left corner indicates higher accuracy.}   
   \label{fig:timestamp_accuracy}
\end{figure}


\Cref{fig:timestamp_accuracy} shows the word-level timestamp estimation accuracy as a function of an estimation error tolerance threshold. For each x-axis value, the corresponding y-axis value shows the percentage of words whose estimated timestamp falls within this threshold when compared to the reference timestamp. A curve that approaches the upper-left corner indicates a model with more accurate timestamp estimation. Universal-1 model produced more accurate timestamps than Whisper large-v3 by a significant margin. Additionally, we included two powerful open-source models, NeMo's Parakeet RNN-T and CTC models with 600M parameters. The results highlight the advantage of RNN-T in terms of timestamp estimation, while our model still substantially outperformed Parakeet RNN-T.

The superior performance achieved by our model is intriguing because an RNN-T loss does not explicitly include any terms related to the accuracy of audio-to-token alignments~\citep{zhao2021addressing}. Nevertheless, a few observations can be made to describe this unexpected effectiveness. First, the use of bidirectional context by the encoder is essential. In our other experiments using a streaming Conformer encoder, the timestamp prediction accuracy dropped drastically, with significant delays compared to the reference timestamps. In the streaming scenario, an additional mechanism, such as the one proposed in \citet{zhao2021addressing}, is necessary to compensate for the delays. Second, greedy decoding may be more beneficial for timestamp accuracy than beam search, which optimizes for transcription accuracy. Finally, the use of chunk-wise attention in the encoder may also help constrain the audio-to-token alignment space. Further research into these aspects will allow us to draw more decisive conclusions.

\begin{figure}[ht]
   \centering
   \includegraphics[width=0.5\textwidth]{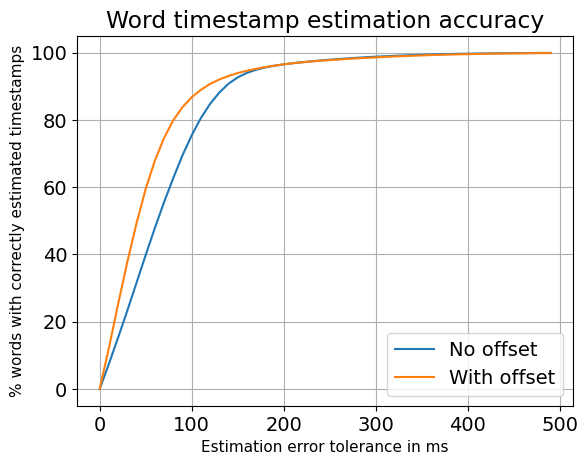} 
   \caption{Effect of bias offset in word-level timestamp estimation.}
   \label{fig:timestamp_offset}
\end{figure}

A constant offset of -65ms was applied in our implementation as we found that the raw timestamp estimates generated by our model were slightly biased across test samples. The offsetting constant was obtained as the median value of the biases computed over a subset of the data. \Cref{fig:timestamp_offset} shows the effect of the bias offset.

\subsection{Impact of Pre-training} \label{sec:pretraining-impact}

\begin{figure}[ht]
   \centering
   \includegraphics[width=0.6\textwidth]{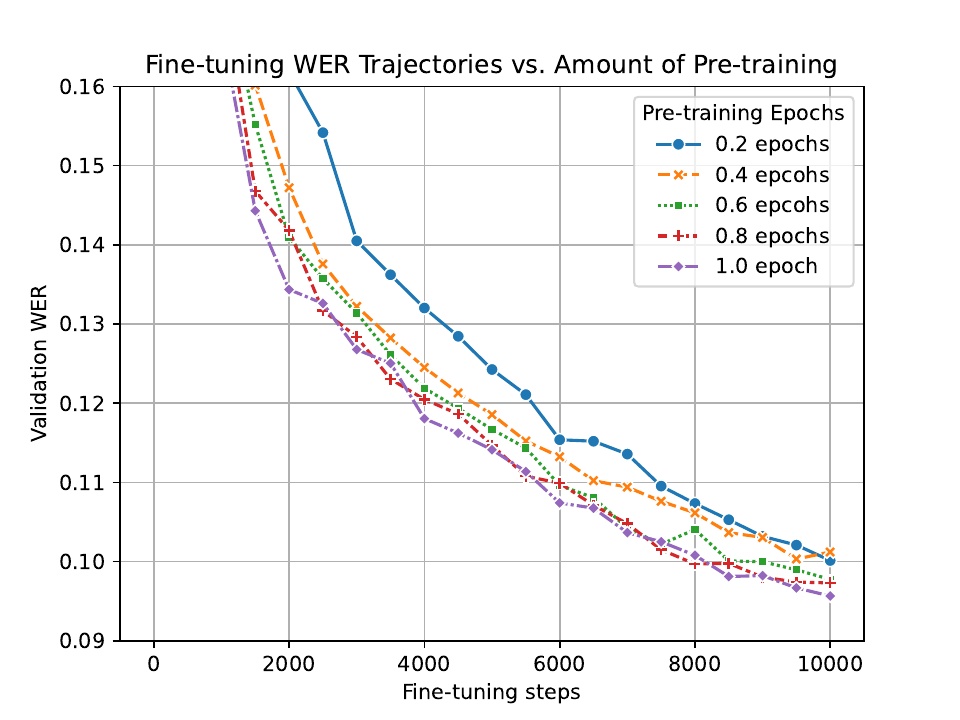} 
   \caption{Fine-tuning WER trajectory when initialized from various pre-training checkpoints}
   \label{fig:pretraining-impact}
\end{figure}

Finally, we studied the impact of the amount of pre-training data on the learning trajectory of the fine-tuning stage. To do so, we trained CTC models starting from different pre-training checkpoints. For efficiency in the investigation, for each pre-trained checkpoint, we used the first 12 layers of the encoder, added a CTC decoder on top, and fine-tuned the model for 10k steps to assess performance and convergence speed. Thus, each model size was approximately 300M parameters.

\Cref{fig:pretraining-impact} shows the learning curves of the individual training runs, each starting from a distinct pre-training checkpoint. The validation set consists of 1024 randomly chosen utterances from LibriSpeech and MLS test sets. We can observe that longer BEST-RQ training considerably improves the WER, with diminishing returns around 0.8 epochs on our data, corresponding to 10M hours of audio. Further test WERs are provided in Table \ref{tab:pretrain_ablation}, showing a similar performance trend. We also note that, without pre-training, our end-to-end ASR training did not converge. This indicates the usefulness of pre-training in stabilizing training models without extensive hyperparameter tuning.


\begin{table}[t]
    \centering
    \caption{WERs (\%) after 10k fine-tuning steps, starting from different pre-training checkpoints}.    
    \label{tab:pretrain_ablation}    
    \vspace{-.5em}
    \begin{tabular}{lccccc}
    \toprule
 & \multicolumn{5}{c}{Pre-training epochs} \\
 & {0.2} & {0.4} & {0.6} & {0.8} & {1} \\
\midrule
{Podcast} & 13.0 & 12.4 & 12.3 & 12.1 & 12.2 \\
{Noisy} & 16.5 & 15.6 & 15.4 & 14.9 & 15.0 \\
{LibriSpeech test-clean}  & 4.2 & 4.1 & 4.1 & 4.1 & 4.0 \\
{LibriSpeech test-other} & 9.3 & 8.9 & 8.8 & 8.5 & 8.5 \\
{MLS Spanish}  & 7.9 & 7.8 & 7.6 & 7.7 & 7.5 \\
{MLS French}  & 9.7 & 9.6 & 9.4 & 9.4 & 9.4 \\
{MLS German}  & 10.5 & 10.3 & 10.1 & 10.1 & 10.0 \\
\bottomrule
\end{tabular}
\end{table}

\section{Conclusion}

In this comprehensive study, we have presented Universal-1, a highly capable multilingual ASR system developed for English, Spanish, German, and French. Our investigation has revealed that the combination of a vast pre-training dataset, meticulously curated fine-tuning corpora, and the employment of a Conformer RNN-T model architecture results in remarkably robust performance across languages and test sets while achieving significantly faster inference than state-of-the-art open-source models. 
This achievement underscores the effectiveness of scaling in ASR systems, aligning with the prevailing trend in AI research that emphasizes the importance of the quality and volume of the training data. 

Moreover, our analysis extends beyond a WER comparison, shedding light on practically relevant aspects, such as code-switching, resistance to hallucinations, timestamp estimation, and inference latency, as well as the impact of pre-training on fine-tuning efficiency. The insights derived from these observations could offer the speech community a deeper understanding of the nuanced dynamics within ASR systems.

In this work, we have emphasized a holistic, system-centric approach to the design and analysis of ASR systems. Rather than isolating individual components and investigating them by normalizing other variables, our methodology considers the ASR system in its entirety, focusing on real world use cases and aiming to make more nuanced and practically relevant observations. We believe this perspective is crucial for advancing the field, as the ASR technology matures and finds more applications in the real world. 
Future research should aim to build publicly accessible, comprehensive benchmarks covering the aspects that we discussed in this paper as well as other critical aspects, such as measuring proper noun recognition. 


\bibliographystyle{plainnat}
\bibliography{refs}

\newpage
\appendix
\section{Pre-training details} \label{app:pre-training}
In this section, we provide additional details about model pre-training and encountered issues.

\subsection{Model initialization}
For pre-training, we initialize all linear and convolutional layer weights with a Kaiming uniform distribution, and linear biases with a uniform distribution.

\subsection{Training divergence and stability}
When scaling the pre-training to larger model sizes, especially exceeding 1B parameters, we observed sudden divergence. These divergent model training runs are characterized by sudden loss spikes and a collapse of the distribution of predicted labels. Following the methodology in \citet{wortsman2023smallscale}, we were able to identify the AdamW $\epsilon$ value as the root cause of this issue. We could resume previously diverged training runs by lowering the $\epsilon$ value from $10^{-8}$ to $10^{-15}$.

In addition to the above, we observed more stable RNN-T training when starting from a pre-trained checkpoint. In other words, training runs that would diverge when training from scratch will converge when initializing the model from a pre-trained checkpoint. More details can be found in \Cref{sec:pretraining-impact}.

\section{Evaluation Datasets} \label{app:eval_data}
\subsection{English datasets}
\begin{enumerate}
    \item Common Voice V5.1: We used the English subset of the V5.1 dataset from the official website.
\item CORAAL: We used the version 2021.07 dataset from official sources and segmented according to the FairSpeech project.
\item TED-LIUM 3: We used 11 TED talks, following the Whisper’s TED-LIUM 3 long-form partition.
\item LibriSpeech: We used the test-clean and test-other splits from the LibriSpeech ASR corpus.
\item Earnings-21: We used the corpus of earnings calls from the speech-datasets repository from the 202206 version.
\item Meanwhile: We followed the dataset creation procedure from Whisper and downloaded the 64 segments from YouTube.
\item Podcast: We used a 18.2-hour human-labeled dataset of podcasts.
\item Broadcast: We used a 7.5-hour human-labeled private dataset of news broadcasts.
\item Telephony: We used a 8.2-hour human-labeled private dataset of telephone conversations.
\item Noisy: We used a 7.2-hour human-labeled private dataset of noisy real world audio.
\end{enumerate}
\subsection{Multilingual datasets}
\begin{enumerate}

\item Fleurs: We downloaded the test splits for each language from the HuggingFace distribution.
\item MLS: We used the test split of each language in the Multilingual LibriSpeech (MLS) corpus.
\item VoxPopuli: We downloaded and segmented the dataset according to the official instructions for each language.
\item Common Voice V9: We downloaded the V9 dataset from the official website for each language.
\item Private: We used a small 6--10 hour human-labeled dataset of real-world audio.
\end{enumerate}

\subsection{Timestamp estimation datasets}\label{app:eval_data_timestamps}
\begin{enumerate}

\item LibriSpeech test: We merge both test-clean and test-other of the LibriSpeech ASR corpus. 
\item Private: We use an internal test set consisting of audio across 4 audio domains (Telephony, Podcasts, Broadcast, and Webinars).
\item Challenge: We used a more difficult test set containing targeted real-world edge cases, such as long silences and multi-talker conversations, in addition to a subset of LibriSpeech and Private.
\end{enumerate}

\section{Text formatting} \label{app:text_formatting}


The output from an ASR system typically consists of a sequence of words with no formatting (e.g., \textit{how much is the new phone you got i think something around five hundred dollars}). Even when the ASR output text represents the spoken words accurately, the absence of proper punctuation, truecasing, and other formatting artifacts makes the text challenging for humans to read. Therefore, in real-world applications, we apply a post-processing step to convert the ASR output into a formatted version that is more suitable for human readers (e.g., \textit{How much is the new phone you got? I think something around \$500.}).



To achieve this, we employ a multi-step pipeline approach that takes the raw ASR output and formats it into written text by applying the following processing steps. 
\begin{enumerate}
    \item \textbf{Inverse Text Normalization}: This step converts certain spoken-format phrases into its written form  (e.g., \textit{five hundred dollars} to \textit{\$500}). 
    \item \textbf{Punctuation Restoration}: This step adds punctuation marks, such as periods, commas, and question marks.
    \item \textbf{Truecasing}: This step capitalizes letters where appropriate, such as the starts of sentences and proper nouns. 
\end{enumerate}

\end{document}